\documentclass[11pt]{article}
\usepackage{amssymb}
\usepackage{url}
\usepackage{a4}
\usepackage{cite}
\usepackage{ntheorem}
\usepackage{graphicx}  

\usepackage{epsfig}
\usepackage{xcolor}

{\catcode `\@=11 \global\let\AddToReset=\@addtoreset}
\AddToReset{equation}{section}
\renewcommand{\theequation}{\thesection.\arabic{equation}}

\newcounter{mnotecount}[section]

\newcommand{\rmnote}[1]{}

\newcommand{\cH}{{\cal H}}
\newcommand{\OOmega}{\Omega}

\newcommand{\riemgz}{g_0}

\renewcommand{\hbar}{{\overline \riemgz}}

\newcommand{\letters}

\newcommand{\mcE}{{\mycal E}}

\newcommand{\mcW}{{\mycal W}}

\newcommand{\nablash}{\nabla{\kern -.75 em
     \raise 1.5 true pt\hbox{{\bf/}}}\kern +.1 em}
\newcommand{\Deltash}{\Delta{\kern -.69 em
     \raise .2 true pt\hbox{{\bf/}}}\kern +.1 em}
\newcommand{\Rslash}{R{\kern -.60 em
     \raise 1.5 true pt\hbox{{\bf/}}}\kern +.1 em}

\newcommand{\hyp}
{{\mycal S}}

\newcommand{\mcM}{{\mycal M}}

\newcommand{\bea}{\begin{eqnarray}}
\newcommand{\beaa}{\begin{eqnarray*}}
\newcommand{\bean}{\begin{eqnarray}\nonumber}

\newcommand{\bel}[1]{\begin{equation}\label{#1}}
\newcommand{\beal}[1]{\begin{eqnarray}\label{#1}}
\newcommand{\beadl}[1]{\begin{deqarr}\label{#1}}
\newcommand{\eeadl}[1]{\arrlabel{#1}\end{deqarr}}
\newcommand{\eeal}[1]{\label{#1}\end{eqnarray}}
\newcommand{\eead}[1]{\end{deqarr}}
\newcommand{\eea}{\end{eqnarray}}
\newcommand{\eeaa}{\end{eqnarray*}}

\newcommand{\be}{\begin{equation}}
\newcommand{\ee}{\end{equation}}

\newcommand{\myeq}[1]{(\ref{#1})}

\DeclareFontFamily{OT1}{rsfs}{}
\DeclareFontShape{OT1}{rsfs}{m}{n}{ <-7> rsfs5 <7-10> rsfs7 <10->
rsfs10}{} \DeclareMathAlphabet{\mycal}{OT1}{rsfs}{m}{n}
\def\scri{{\mycal I}}%
\def\scrip{\scri^{+}}%
\def\Scri{\scri}

\theorembodyfont{\upshape}

\def \Reel{\mathbb{R}}

\def \R {\Reel}

\newcommand{\xflat}{X^\flat}
\newcommand{\kflat}{K^\flat}
\newcommand{\yflat}{Y^\flat}
\newcommand{\Sext}{\hyp_{\mathrm ext}}
\newcommand{\doc}{\langle\langle \mcM\rangle\rangle}

\newcommand{\tcM}{\,\,\,\,\widetilde{\!\!\!\!\cM}}
\newcommand{\tg}{{\tilde g}}

\newcommand{\cg}{\,{\tilde {\!g}}}

\newcommand{\cM}{\mycal M}

\newcommand{\scrim}{\scri^-}

\newcommand{\mcB}{{\mycal B}}

\begin{document}

\title{Stationary Black Holes}

\author{Robert Beig\thanks{Faculty of Physics, University of Vienna, Boltzmanngasse 5, A-1090 Vienna, Austria}
\thanks{Email \protect\url{robert.beig@univie.ac.at}}
\ 
and 
 Piotr T. Chru\'{s}ciel$^*$\thanks{Email \protect\url{piotr.chrusciel@univie.ac.at}, 
 URL
\protect\url{https://homepage.univie.ac.at/piotr.chrusciel} 
}
}
\date{\today}

\maketitle


\begin{abstract}
We review the theory of uniqueness of
 stationary black hole solutions of vacuum Einstein equations.
\end{abstract}


\section{Introduction}\label{Sin}

In this article we consider a specific class of stationary solutions to
the Einstein field equations, which read
\begin{equation}
\label{einstein} R_{\mu\nu} - \frac{1}{2}g_{\mu\nu}R = \frac{8 \pi
G}{c^{4}} T_{\mu\nu}.
\end{equation}
Here $R_{\mu\nu}$ and $R=g^{\mu\nu}R_{\mu\nu}$ are respectively
the Ricci tensor and the Ricci scalar  of the spacetime metric
$g_{\mu\nu}$, $G$ is the Newton constant and $c$ the speed of
light. The tensor $T_{\mu\nu}$ is the stress-energy tensor of
matter. Spacetimes, or regions thereof, where $T_{\mu\nu}=0$ are
called vacuum.

 Stationary solutions are of interest for a variety of reasons. As models for
compact objects at rest, or in steady rotation, they play a key
role in astrophysics. They are easier to study than non-stationary
systems because stationary solutions are governed by elliptic
rather than hyperbolic equations. Finally, like in any field
theory, one expects that large classes of dynamical solutions
approach (``settle down to") a stationary state in the final
stages of their evolution.

The simplest stationary solutions describing compact isolated
objects are the spherically symmetric ones. In the vacuum region
these are all given by the Schwarzschild family. A theorem of Jebsen, 
usually attributed to
Birkhoff, shows that in the vacuum region any spherically symmetric
metric, even without assuming stationarity, belongs to the family
of Schwarzschild metrics, parameterized by a mass
parameter $m$. 
Thus, regardless of possible motions of the matter, as long as
they remain spherically symmetric, the exterior metric is the
Schwarzschild one with some constant $m$, which for usual 
classical-matter models is positive. This has the following
consequence for stellar dynamics: Imagine following the collapse
of a cloud of pressureless fluid (``dust"). Within Newtonian
gravity this dust cloud will, after finite time, contract to a
point at which the density and the gravitational potential
diverge. However, this result cannot be trusted as a sensible
physical prediction because, even if one supposes that Newtonian
gravity is still valid at very high densities, a matter model
based on non-interacting point particles is certainly not.
Consider, next, the same situation in the Einstein theory of
gravity: Here a new question arises, related to the form of the
Schwarzschild metric outside of the spherically symmetric body:
\begin{equation}
 g=-V^{2}dt^{2}+V^{-2}dr^{2}+r^{2}d\Omega^{2}\;,
\quad V^{2}=1-\frac{2Gm}{rc^{2}} \;,\quad t\in \R\;,\ r\in
 (\frac{2Gm}{c^{2}},\infty)
 \,.
 \label{Schwarz} 
\end{equation} 
Here $d\Omega^{2}$ is the line element of the standard
2-sphere. When $m>0$ the metric \myeq{Schwarz} is manifestly singular as
$r=2m  
$ 
is approached (from now on we use units in which $G=c=1$), and there
arises the need to understand what happens
when the radius $r=2m$  
is reached, and crossed.

The first key feature of the metric (\ref{Schwarz}) is its
stationarity, with Killing vector field $X$ given by
$X=\partial_t$. A Killing field, by definition, is a vector field
the local flow of which generates isometries. An asymptotically flat
spacetime%
\footnote{We use the term \emph{spacetime} to denote  a smooth,
paracompact, connected, orientable and time--orientable four-dimensional
Lorentzian
manifold.} 
is called {\em stationary } if there exists a Killing
vector field $X$ which approaches $\partial_t$ in the
asymptotically flat region (where $r$ goes to $\infty$, see below
for precise definitions) {\em and} generates a one parameter
groups of isometries.   A spacetime is called {\em static} if it
is stationary and if the stationary Killing vector $X$ is
hypersurface-orthogonal, i.e. $\xflat \wedge d \xflat = 0$, where
$\xflat = X_\mu dx^\mu = g_{\mu\nu} X^\nu dx^\mu$. A spacetime
is called {\em axisymmetric} if there exists a Killing vector
field $Y$, which generates a one parameter group of isometries,
and which behaves like a {\em rotation} in the asymptotically flat
region, with  all orbits $2\pi $-periodic. In asymptotically flat
spacetimes this implies that there exists an axis of symmetry,
that is, a set on which the Killing vector vanishes. Killing
vector fields which are a non-trivial linear combination of a time
translation and of a rotation in the asymptotically flat region
are called \emph{stationary-rotating}, or \emph{helical}.

\begin{figure}[t]
\label{Sfig0}%
\begin{center}
\includegraphics[width=.8\textwidth]{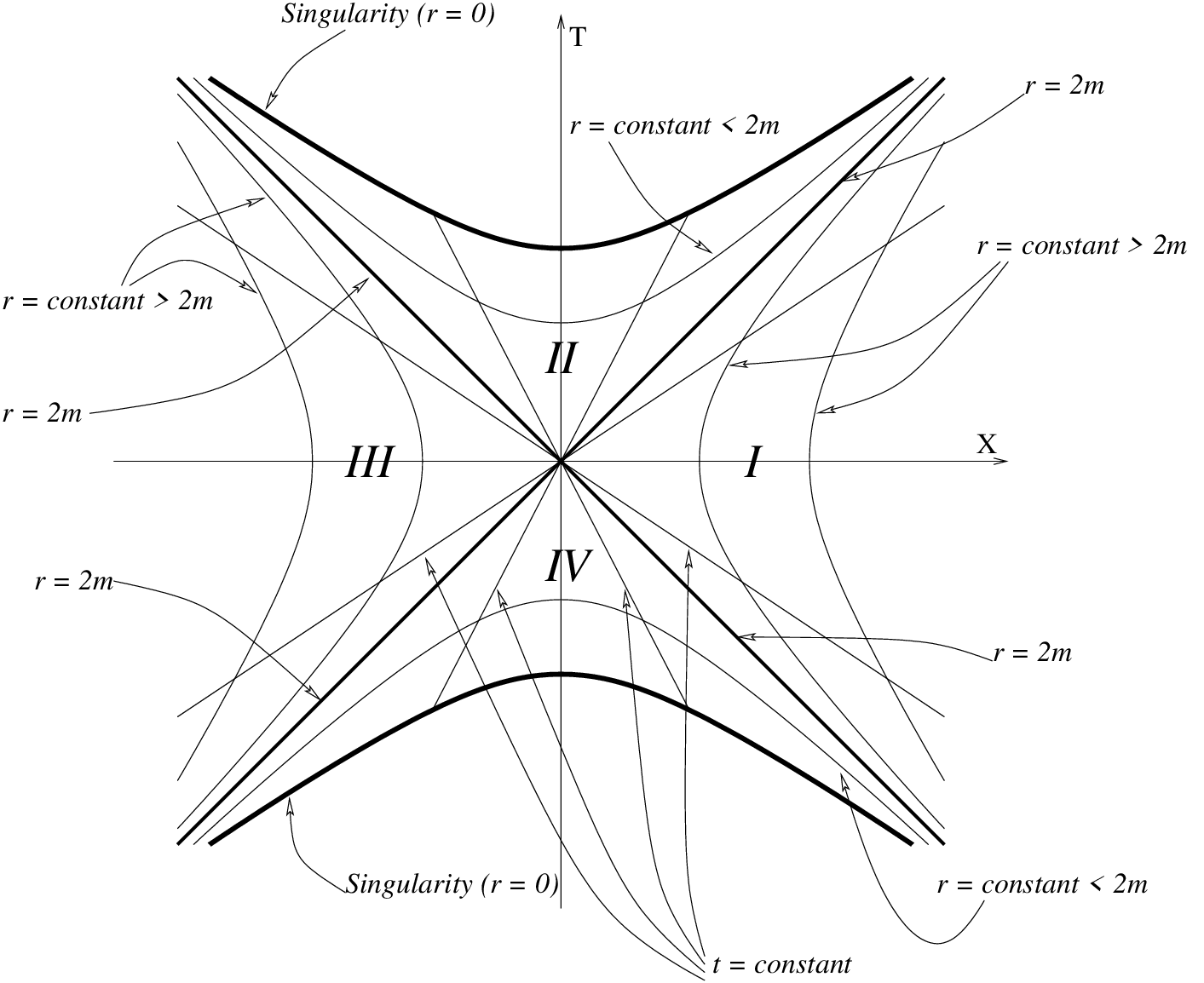}
\end{center}
\caption{The Kruskal-Szekeres extension of the Schwarzschild
solution.}
\end{figure}

There exists a technique, due independently to Kruskal and
Szekeres, of attaching together two regions $r>2m$ and two regions
$r<2m$ of the Schwarzschild metric, in a way
shown\footnote{\label{fNi}We are grateful to J.-P.~Nicolas for
allowing us to use his electronic figures, based on those in
\emph{Dissertationes Math.} {\bf 408} (2002), 1--85.} in
Figure~\ref{Sfig0}, to obtain a manifold with a metric which is
smooth at  $r=2m$. In the extended spacetime the hypersurface
$\{r=2m\}$ is a null hypersurface $\mcE$, the Schwarzschild event
horizon. The stationary Killing vector $X=\partial_t$ extends to a
Killing vector in the extended spacetime, and becomes tangent to
and null on $\mcE$.
 The global properties of the Kruskal--Szekeres
extension of the exterior Schwarzschild%
\footnote{The reader is warned that the Kruskal--Szekeres extension is only one out of many possibilities. Indeed, the exterior
Schwarzschild spacetime (\ref{Schwarz}) admits an infinite number of
non-isometric vacuum extensions, even in the class of maximal,
analytic, simply connected ones. The Kruskal-Szekeres extension is
singled out by the property that it is vacuum,
analytic, simply connected, with   the area of the orbits of the
isometry group tending to zero along incomplete geodesics.} 
spacetime make this
spacetime a natural model for a non-rotating black
hole.

We can now come back to the question of the contracting dust cloud
according to the Einstein theory. For simplicity we take the
density of the dust to be uniform --- the so-called
Oppenheimer--Snyder solution. It then turns out that, in the course
of collapse, the surface of the dust will eventually cross the
Schwarzschild radius, leaving behind a Schwarzschild black hole.
If one follows the dust cloud further, a singularity will
form, but will not be visible from the "outside world",
where $r>2m$. For solar-mass objects the Schwarzschild radius
is of the order of 
kilometers, and it is of the order of millions of kilometers for the black hole  in the center of our galaxy, so
standard
phenomenological matter models such as that for dust can still be
trusted, and the previous objection to the Newtonian scenario does
not apply.

There is a rotating generalization of the Schwarzschild metric,
namely the two parameter family of \emph{exterior Kerr metrics},
which in Boyer-Lindquist coordinates take the form
\begin{eqnarray}
\label{Kerr} \lefteqn{ g = -\frac{\Delta - a^{2} \sin^{2}\theta
}{\Sigma}dt^{2} - \frac{2a \sin ^{2} \theta
(r^{2}+a^{2}-\Delta)}{\Sigma}dtd\varphi  + {} }
\nonumber\\
& & {}+ \frac{(r^{2}+a^{2})^{2}-\Delta a^{2}\sin ^{2}\theta
}{\Sigma}\sin ^{2}\theta  d\varphi
^{2}+\frac{\Sigma}{\Delta}dr^{2} +\Sigma d\theta ^{2},
\end{eqnarray}
with $0 \leq a<m$. Here $\Sigma=r^{2}+a^{2}\cos^{2}\theta $,
$\Delta=r^{2}+a^{2}-2mr$ and $r_{+}<r<\infty$ where
$r_{+}=m+(m^{2}-a^{2})^{\frac{1}{2}}$. When $a=0$, the Kerr metric
reduces to the Schwarzschild metric (though this is not obvious in the coordinates of (\ref{Kerr})). 
The Kerr metric is again a
vacuum solution, and it is stationary with $X=\partial_t$, the
asymptotic time translation, as well as axisymmetric with
$Y=\partial_\varphi$, the generator of rotations. Similarly to the
Schwarzschild case, it turns out that the metric can be smoothly
extended across $r=r_{+}$, with $\{r=r_{+}\}$ being a smooth null
hypersurface $\mcE$ in a suitable extension. The null generator $K$ of
$\mcE$ is the extension
 of the stationary-rotating Killing field $X +
\OOmega Y$, where $\OOmega = \frac{a}{2mr_{+}}$. On the other
hand, the Killing vector $X$ is timelike only outside the
hypersurface $\{r=m+(m^{2}-a^{2} \cos^{2}\theta )^\frac{1}{2}\}$,
on which $X$ becomes null. In the region between $r_+$ and
$r=m+(m^{2}-a^{2}\cos^{2}\theta )^\frac{1}{2}$, which is called
the \emph{ergoregion}, $X$ is spacelike. It is also spacelike on
and tangent to  $\mcE$, except where the axis of rotation meets
$\mcE$, where $X$ is null. By the above properties the Kerr family
provides natural models for rotating black holes.

Unfortunately, as opposed to the spherically symmetric case, there
are no known explicit collapsing solutions with rotating matter,
in particular no known solutions having the Kerr metric as final
state.

The aim of the theory outlined below is to understand the general
geometrical features of stationary black holes, and to give a
classification of models satisfying the field
equations.

\section{Model independent concepts}

We now make precise some notions used informally in the
introductory section.  The mathematical notion of black hole is
meant to capture the idea of a region of spacetime which cannot
be seen by ``outside observers". Thus, at the outset, one assumes
that there exists a family of physically preferred observers in
the spacetime under consideration. When considering isolated
physical systems, it is natural to define the ``exterior
observers" as observers which are ``very far" away from the system
under consideration.
 The standard way of making this mathematically precise is by using conformal completions,  discussed in
more detail in the article about asymptotic structure in this
encyclopedia: A pair $(\tcM ,\cg)$ is called a \emph{conformal
completion at infinity,} or simply {\em conformal completion}, of
$(\mcM,g)$ if $\tcM $ is a manifold with boundary such that:
\begin{enumerate}\item $\mcM$ is the interior of $\tcM $, \item there exists a function $\Omega$, with the
property that the metric $\tg$, defined as $\Omega^{2}g$ on
$\mcM$, extends by continuity to the boundary of $\tcM$, with the
extended metric remaining of Lorentzian signature, \item $\Omega$
is positive on $\mcM$, differentiable on $\tcM $, vanishes on the
boundary $$\Scri:=\tcM\setminus\mcM\;,$$ with $d\Omega$
\emph{nowhere vanishing} on $\Scri$.
\end{enumerate} The boundary $\scri$ of~$\tcM $ is called Scri, a phonic shortcut for ``script I".
 The idea here is the following: forcing $\Omega$ to vanish on
 $\Scri$ ensures that $\Scri$ lies infinitely far away from any
 physical object ---  a mathematical way of capturing the notion ``very far
 away". The condition that $d\Omega$ does not vanish
 is a convenient technical condition which ensures that $\Scri$ is a smooth
 three dimensional hypersurface, instead of some, say, one or two dimensional
 object, or of a set with singularities here and there. Thus,
 $\Scri$ is an idealized description of a family of observers at
 infinity.

To distinguish between various points of $\scri$ one sets
$$
\Scri^+=\{\mbox{points in $\Scri$ which are to the timelike future of the
physical spacetime}\}\ .
$$$$
\Scri^-=\{\mbox{points in $\Scri$ which are to the timelike past of the
physical spacetime}\}\ .
$$
(Recall that a point $q$ is to the timelike future, respectively 
past, of $p$ if there exists a future directed, respectively past
directed, timelike curve from $q$ to $p$.  Timelike curves are curves
$\gamma$ such that their tangent vector $\dot \gamma$ is timelike
everywhere, $g(\dot \gamma,\dot \gamma)< 0$. The set of points related by a future directed timelike curve with a set $\Omega$ is denoted by $I^+(\Omega)$. Similarly a curve is causal if $\dot \gamma$ is nowhere vanishing, with $g(\dot \gamma,\dot \gamma)\le  0$ everywhere.) 
 One then defines the black hole
region $\mcB$ as 
\bean \mcB &:=&  \{\mbox{the set of points in
$\mcM$ from which}
\\ 
&& \phantom{ the}
\mbox {no future directed causal curve in $\tcM$ meets
$\scrip$}\}\;.
\eeal{sbh1}
By definition, points in the black hole
region cannot send information to $\scrip$; equivalently,
observers on $\scrip$ cannot see points in $\mcB$.
 The \emph{white hole} region $\mcW$ is
defined by changing the time orientation in \myeq{sbh1}.
%

 A key notion related to the concept of a black hole is
that of  \emph{future} ($\mcE^+)$ and \emph{past ($\mcE^-$) event
horizons}, 
\bel{eh1} \mcE^+:= \partial \mcB\;,\qquad \mcE^-:=
\partial \mcW\;.
\ee 
Under assumptions spelled-out below,
 event horizons in stationary spacetimes
with matter satisfying the  \emph{null energy condition},
\begin{equation}
T_{\mu \nu }\ell^\mu \ell^\nu \ge 0\quad \mbox{  for all null
vectors $\ell^\mu$,} \label{energy}
\end{equation} are  {smooth} null
hypersurfaces, analytic if the metric is analytic. 

 Indeed, in order to
develop a reasonable theory one also needs  regularity conditions
for the interior of spacetime.   These have to be  conditions
that do not exclude singularities (otherwise the Schwarzschild
and Kerr black holes would be excluded), but which nevertheless
guarantee a well-behaved exterior region. One such condition,
assumed in all the results described below, is the existence in
$\mcM$ of an asymptotically flat space-like hypersurface $\hyp$
which is the union of an asymptotic regions diffeomorphic to $\R^3$ minus a ball, and of a compact set.
One further assumes that either $\hyp$ has no boundary, or
the boundary of $\hyp$ lies on $\overline{\mcE^+\cup\mcE^-}$. To
make things precise, for any spacelike hypersurface let $g_{ij}$
be the induced metric, and let $K_{ij}$ denote its extrinsic
curvature.
 A space--like hypersurface
$\hyp_{\mathrm{ext}}$ diffeomorphic to
      ${\Bbb R}^3$ minus a
ball will be called \emph{asymptotically flat} if the fields
$(g_{ij},K_{ij})$  satisfy the fall--off conditions
\begin{equation}
|g_{ij}-\delta _{ij}|+r|\partial _\ell g_{ij}|+\cdots
+r^k|\partial _{\ell _1\cdots \ell _k}g_{ij}|+r|K_{ij}|+\cdots
+r^k|\partial _{\ell _1\cdots \ell _{k-1}}K_{ij}|\le Cr^{-1 }\ ,
\label{falloff}
\end{equation}
for some constants $C$,  $k\ge 1$. A hypersurface $ \hyp $ (with
or without boundary) will be said to be {\em asymptotically flat
with compact interior} if $\hyp $ is of the form $\hyp_{int}\cup
\hyp_{\mathrm{ext}}$, with $\hyp_{int}$ compact and $\Sext$
asymptotically flat.

There exists a canonical way of constructing a conformal
completion with good global properties for stationary spacetimes
which are asymptotically flat in the sense of \myeq{falloff}, and
which are vacuum sufficiently far out in the asymptotic region.
This conformal completion is referred to as the \emph{standard
completion}
and will be assumed from now on.

Given the above completion, the \emph{domain of outer communications} (d.o.c.) of a black hole spacetime is defined as
  \bel{dext} \doc:= \mcM \setminus\{\mcB\cup\mcW\}\;.\ee
  Thus, $\doc$ is the region lying outside of the white hole region
  and outside of the black hole region; it is the region which can
  both
  be seen by the outside observers and  influenced by
  those.
  A natural causal regularity condition, which prevents things like existence of closed timelike curves, is to require that $\doc$ be \emph{globally hyperbolic}, i.e., there exists a spacelike hypersurface which meets every inextendible timelike curve precisely once. This could be, but does not have to, be the hypersurface $\hyp$ above.

Yet another useful condition is  that there exists a cross-section of $\scrim$, say $K$,   so that the boundary $\dot I^+(K)$ of the timelike future of $K$ intersects the future event horizon $\mcE^+$ in a compact crosssection. 

The collection of all regularity conditions spelled-out so far is
known as \emph{$I^+$-regularity}, and it will be assumed in everything that follows. 

\begin{figure}[t]
\begin{center}
  \includegraphics[width=.8\textwidth]{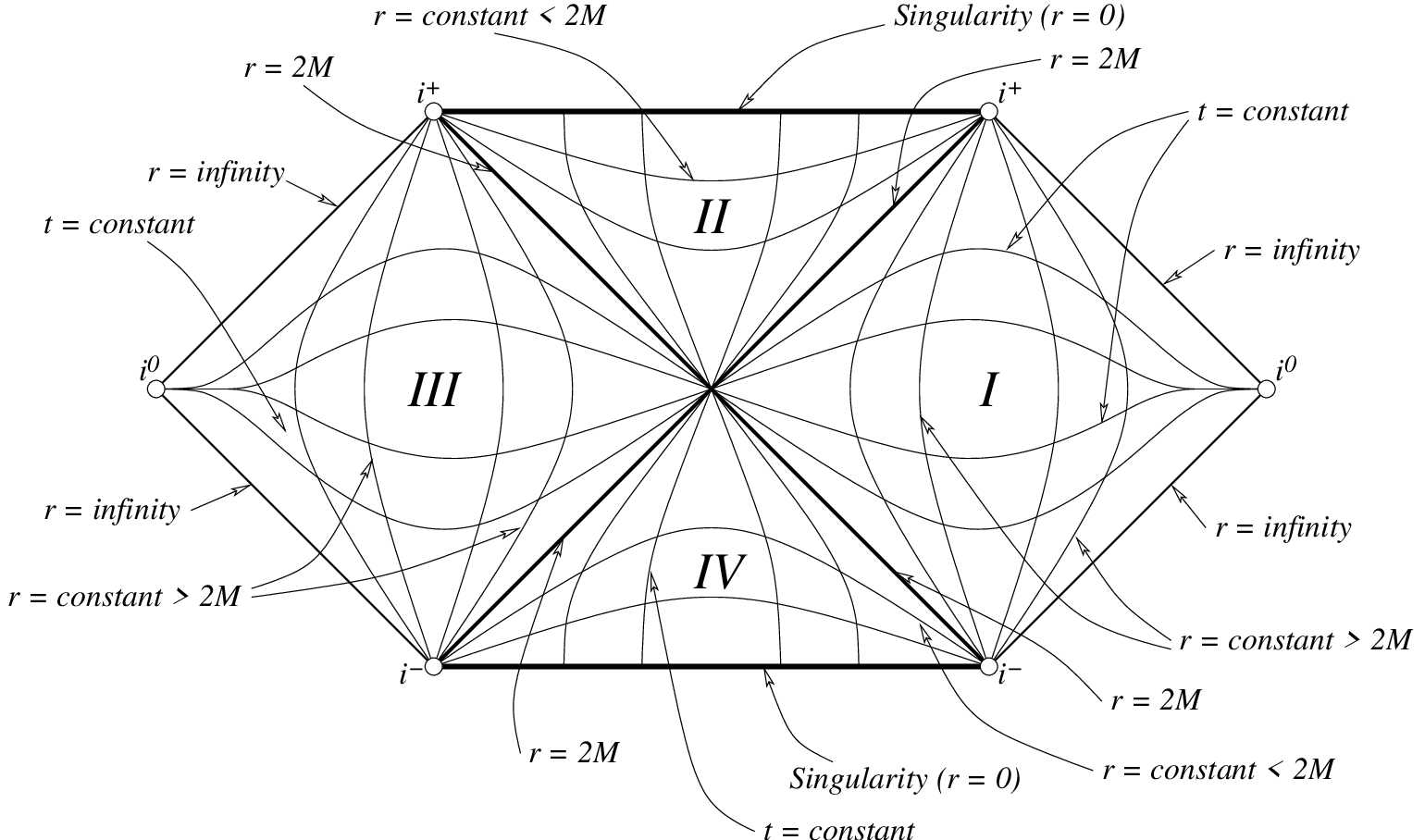}
\end{center}
\caption[FCPd]{The Carter-Penrose diagram for the Kruskal-Szekeres
spacetime. The spacetime contains two asymptotically flat regions, regions $I$ and $III$ in the figure,
with corresponding $\scri^\pm$ and $\mcE^\pm$ defined with respect
to the second region. Each
point in this diagram represents a two-dimensional sphere, and
coordinates are chosen so that light-cones have slopes plus minus
one. Regions are numbered as in Figure~\ref{Sfig0}.} \label{FCPd}
\end{figure}

Returning to the future event horizon,   it is now fairly straightforward
 to show that every Killing vector field $X$ is 
necessarily tangent to $\mcE^+$. 
As already mentioned, with a considerable amount of work one also proves 
that $\mcE^+$ is a smooth hypersurface. It follows that $X$ is either null or
spacelike on $\mcE^+$. This leads to a preferred class of event
horizons, called \emph{Killing horizons}. By definition, a Killing
horizon associated with a Killing vector $K$ is a \emph{null
hypersurface} which coincides with a connected component of the
set \bel{defKH} \cH(K):=\{p\in\mcM:\ g(K,K)(p)=0\;,\ K(p)\ne
0\}\;.\ee A simple example is provided by the ``boost Killing
vector field" $K=z\partial_t+ t\partial_z$ in Minkowski
spacetime: in this case $\cH(K)$ has four connected components
$$\cH_{\epsilon \delta} :=\{ t=\epsilon z\;, \delta t >0\}\;,
\quad \epsilon, \delta \in \{\pm 1\}\;.$$ The closure $\overline
\cH$ of $ \cH$ is the set $\{|t|=|z|\}$, which is not a manifold,
because of the crossing of the null hyperplanes $\{t=\pm z\}$ at
$t=z=0$. Horizons of this type are referred to as {\em bifurcate
Killing horizons}, with the set $\{K(p)=0\}$ called the
\emph{bifurcation surface} of $\cH(K)$. The bifurcate horizon
structure in the Kruskal-Szekeres extension of the Schwarzschild spacetime can be
clearly seen on Figures~\ref{Sfig0} and \ref{FCPd}.

A careful revisit of  a lemma of Vishveshwara and Carter 
shows that if a Killing vector $K$ is hypersurface-orthogonal,
$\kflat \wedge d \kflat=0$, then the set $\cH(K)$ defined in
\myeq{defKH} is a union of smooth null hypersurfaces, with $K$ being
tangent to the null geodesics threading $\cH$ (``$\cH$ is
generated by $K$"), and so is indeed a Killing horizon. It has
been shown by Carter that the same conclusion can be reached if
the hypothesis of hypersurface-orthogonality is replaced by that
of existence of two linearly independent Killing vector fields.

 In stationary-axisymmetric spacetimes a
Killing vector $K$ \emph{tangent to the generators} of a Killing
horizon $\cal H$ can be normalised so that
 $K=X+\OOmega  Y$,
 where $X$ is the Killing vector field which asymptotes
 to a time translation in the asymptotic region, and $Y$ is the
 Killing vector field which generates rotations in the asymptotic
 region. The constant $\OOmega $ is called the \emph{angular
 velocity of the Killing horizon $\cal H$}.

On a Killing horizon $\cH(K)$ one necessarily has \bel{kappadef}
\nabla^\mu(K^\nu K_\nu)=-2\kappa K^\mu . \ee Assuming the
so-called dominant energy condition on $T_{\mu\nu}$,\footnote{See
the article by Bray on "Positive Energy Theorem and other
inequalities in General Relativity"} it can be shown that $\kappa$
is constant (recall that Killing horizons are always connected in
our terminology), it is called {\em the surface gravity of $\cH$}.
A Killing horizon is called \emph{degenerate} or \emph{extreme} when $\kappa= 0$,
and non--degenerate otherwise; by an abuse of terminology one
similarly talks of degenerate black holes, \emph{etc}. In Kerr
spacetimes we have $\kappa=0$ if and only if $m=a$. 

  The subset of $\doc$ 
 where $X$ is spacelike is called the
 \emph{ergoregion}. In the Schwarzschild spacetime we have $\OOmega =0$ and
 the ergoregion is empty, but   in Kerr with
 $a \ne 0$ there is always a nontrivial ergoregion, enclosed by an \emph{ergosphere}.

 A very convenient method for visualising
  the global structure of spherically symmetric spacetimes is provided by the
  \emph{conformal Carter-Penrose diagrams}. An example of such a diagram is
  provided by Figure~\ref{FCPd}.
  Without spherical symmetry such diagrams can sometimes be replaced   by \emph{projection diagrams}, which 
  give a reasonably accurate causal representation of e.g. Kerr metrics.

  A corollary of the \emph{topological censorship theorem} of
Friedman, Schleich and Witt is that d.o.c.'s of regular black hole
spacetimes satisfying the dominant energy condition are simply
connected. This implies that connected components of event
horizons in stationary spacetimes have ${\Bbb R}\times S^{2}$
topology.

We end our review of the concepts associated with stationary black
hole spacetimes by summarising the properties of the Schwarzschild
and Kerr geometries: The standard textbook extension of the Kerr spacetime with $m
> |a|$ is a black hole spacetime with the hypersurface  $\{r =
r_+\}$ forming a non-degenerate, bifurcate Killing horizon
generated by the vector field $X + \OOmega  Y \equiv \partial_t + \OOmega \partial_\varphi$ and surface gravity
given by $\kappa =
\frac{(m^{2}-a^{2})^{1/2}}{2m[m+(m^{2}-a^{2})^{1/2}]}$. In the
case $a=0$, where the angular velocity $\OOmega $ vanishes, $X$ is
hypersurface-orthogonal and becomes the generator of $\cH$. The
bifurcation surface in this case is the totally geodesic 2-sphere,
along which the four regions in Figure 1 are joined.

\section{Classification of stationary solutions (``No hair
  theorems")}

We confine attention to the ``outside region" of black holes, the
domain of outer communications (d.o.c.).\footnote{Except for the
so-called degenerate case discussed later, the ``inside"(black
hole) region is not stationary, so that this restriction already
follows from the requirement of stationarity.} For reasons of
space we only consider vacuum solutions; there exists a similar
theory for electro-vacuum black holes. (There is a somewhat less
developed theory for black hole spacetimes in the presence of
nonabelian gauge fields, see the review by Gal'tsov and Volkov.)
In connection with a collapse scenario the vacuum condition begs
the question: collapse of what? The answer is twofold: First, there
are large classes of solutions of Einstein equations describing
pure gravitational waves. It is believed that sufficiently strong
such solutions will form black holes. (Whether or not they will do
that is related to the \emph{cosmic censorship conjecture},
discussed in the article on Spacetime Topology, Global Structure
and Singularities  in this encyclopedia.) Consider, next,  a
dynamical situation in which matter is initially present. The
conditions imposed in this section correspond then to a final
state in which matter has either been radiated away to infinity,
or has been swallowed by the black hole (as in the spherically
symmetric Oppenheimer--Snyder collapse described above).

Based on the facts below, it is conjectured that the d.o.c.'s of
appropriately regular, stationary, vacuum black holes are
isometrically diffeomorphic to those of  Kerr black holes:

\begin{enumerate}
\item  The \emph{rigidity theorem} (key idea by Hawking): event horizons in
regular, stationary, \emph{analytic}
vacuum black holes are  either {\em Killing horizons},  or
there exists a second Killing vector in $\doc$.

\item The \emph{Killing horizons theorem}  (Sudarsky-Wald):
\emph{non--degenerate} stationary  vacuum black holes such that
the \emph{event horizon is the union of Killing horizons of the stationary Killing vector}
are {\em static}.

\item  Schwarzschild black holes exhaust the family of {\em
static}  regular vacuum black holes (key ideas by Israel and  Bunting --
Masood-ul-Alam).

\item   Kerr black holes
satisfying
\begin{equation}
\label{acondition} m^{2}\ge a^{2} 
\end{equation} 
exhaust the family of  
{\em stationary--axisymmetric,} vacuum, \emph{connected}
 black holes (key ideas by Carter and Robinson).  
\end{enumerate}

The above results are collectively known under the name of
\emph{no hair theorems}, and they have not provided the final
answer to the problem so far.
Indeed, there are no \emph{a priori} reasons
known for the analyticity hypothesis in the rigidity theorem, except for the \emph{non-degenerate slowly-rotating} case settled by Alexakis, Ionescu and Klainerman.

Yet another key open question is that of existence of
\emph{non-connected} regular stationary-axisymmetric vacuum black
holes. Hennig and Neugebauer have shown that all such configurations with two black-hole components are singular.
Regular stationary vacuum configurations with even more components are not expected to exist either. 
 However, there is the following general result of Weinstein:  Let
$\partial\hyp _b$, $b=1,\ldots,N$ be the connected components of
$\partial \hyp $. Let $\xflat=g_{\mu\nu}X^\mu dx^\nu$, where
$X^\mu$ is the Killing vector field which  
approaches time translations in the asymptotically flat region. Similarly set
$\yflat=g_{\mu\nu} Y^\mu dx^\nu$, $Y^\mu$ being the Killing vector
field associated with rotations. On each $\partial\hyp _b$ there
exists a constant $\OOmega _b$  such that the vector $X+\OOmega _b
Y$ is tangent to the generators of the Killing horizon
intersecting $\partial \hyp_b$. The constant $\OOmega _b$ is
called the angular velocity of the associated Killing horizon.
Define
\begin{eqnarray}%
& 
 \displaystyle 
  m_b =-\frac{1}{8\pi}\int_{\partial\hyp _b}*d\xflat \ , &
\label{ma}
\\
\label{lint} & 
 \displaystyle
 L_b=-\frac{1}{4\pi}\int_{\partial\hyp _b}*d\yflat \
. &
\end{eqnarray}
Such integrals are called \emph{Komar integrals}. One usually
thinks of $L_b$ as the angular momentum of each connected
component of the black hole. Set
\begin{equation}
  \label{mua}
\mu_b=m_b -2 \OOmega _b L_b\;.
  \end{equation}
  Weinstein showed that one necessarily has $\mu_b >0$. The problem at hand can be reduced
  to a \emph{harmonic map} equation, also known as the \emph{Ernst equation},
   involving a singular map from $\R^3$ with Euclidean metric $\delta$ to the two-dimensional hyperbolic space.
    Let $r_b >0$,
  $b=1,\ldots,N-1$, be the distance in $\R^3$  along the axis between
  neighboring black holes as measured with respect to the
  (unphysical) metric $\delta$. Weinstein proved that for \emph{non-degenerate} regular black holes
   the inequality \myeq{acondition} holds, and that
  the metric on $\doc$ is determined up to isometry
  by the $3N-1$ parameters
  \begin{equation}
    \label{par}
    (\mu_1,\ldots,\mu_N,L_1,\ldots,L_N,r_1,\ldots,
r_{N-1})
  \end{equation}
just described, with $r_b,\mu_b>0$. These results by Weinstein
contain the no-hair theorem of Carter and Robinson as a special
case.  Weinstein also shows that for every $N\ge 2$ and for every
set of parameters \myeq{par} with $\mu_b,r_b>0$, there exists a
solution of the problem at hand. It is known that for some sets of
parameters \myeq{par} the solutions will have ``strut
singularities'' between some pairs of neighboring black holes, but
the existence of the ``struts'' for all sets of parameters as
above is not known, and is the main open problem  in our
understanding of \emph{stationary and axisymmetric}  vacuum black
holes.  

\bigskip

\noindent {\bf See also:} Asymptotic  Structure and Conformal
Infinity. Black Hole Thermodynamics. Initial Value problem for
Einstein Equations. Positive energy Theorem and other inequalities
in General Relativity. Spacetime Topology, Causal Structure and
Singularities.

\bigskip

\noindent {\bf Suggestions for further reading:}
\cite{BONeillK,Wald:book,CarterlesHouches,HE,ChBlackHoles,CCH,VolkovGaltsov,Herdeiro:2015waa}

\providecommand{\bysame}{\leavevmode\hbox to3em{\hrulefill}\thinspace}
\providecommand{\MR}{\relax\ifhmode\unskip\space\fi MR }
\providecommand{\MRhref}[2]{%
  \href{http://www.ams.org/mathscinet-getitem?mr=#1}{#2}
}
\providecommand{\href}[2]{#2}

\end{document}